\documentclass[twocolumn,showpacs,preprintnumbers,amsmath,amssymb]{revtex4}
\usepackage{graphicx}
\usepackage{dcolumn}
\usepackage{bm}

\begin{document}

\title{Ground state and excitation dynamics in Ag doped helium clusters}

\author{Massimo Mella}\email{Massimo.Mella@unimi.it} 
\affiliation{
Dipartimento di Chimica Fisica ed Elettrochimica,Universita' degli Studi
di Milano, via Golgi 19, 20133 Milano, Italy}
\author{Maria Carola Colombo}
\altaffiliation{Present address: Laboratory of Inorganic Chemistry, 
ETH H\"onggenberg, CH-8093 Z\"urich, Switzerland}
\email{Colombo@inorg.chem.ethz.ch}
\author{Gabriele Morosi}\email{Gabriele.Morosi@uninsubria.it}
\affiliation{Dipartimento di Scienze Chimiche, Fisiche e Matematiche,
Universita' dell'Insubria,
via Lucini 3, 22100 Como, Italy}

\begin{abstract}
We present a quantum Monte Carlo study of the structure and
energetics of silver doped helium clusters AgHe$_n$ for $n$ up to 100. 
Our simulations show the first solvation shell of the Ag atom to be composed
by roughly 20 He atoms, and to possess a structured angular
distribution. Moreover, the electronic 
$^2$P$_{1/2}\leftarrow ^2$S$_{1/2}$ and $^2$P$_{3/2}\leftarrow ^2$S$_{1/2}$ 
electronic transitions
of the embedded silver impurity have been studied as a function of the number of
helium atoms. 
The computed spectra show a redshift for $n\leq 15$ and
an increasing blueshift for larger clusters, a feature attributed to the
effect of the second solvation shell of He atoms.
For the largest cluster, the computed excitation 
spectrum is found in excellent agreement with the ones recorded in
superfluid He clusters and bulk. No signature of the direct formation of 
proposed AgHe$_2$ exciplex is present in the computed spectra of AgHe$_{100}$.
\end{abstract}

\maketitle

Superfluid $^4$He clusters represent a gentle environment where
high resolution spectroscopic studies of atoms, atomic clusters, and molecules 
at low temperature can be carried out \cite{rev_t}.
In such cold and fluid quantum systems many perturbing effects due to the
temperature and solid matrices are absent, allowing therefore for an easier 
interpretation of the experimentally recorded spectra. 
Moreover, their superfluid behavior allows interesting
quantum effects to take place and to be experimentally probed (for instance see
Refs. \cite{miller,whaley_focus}).

Whereas the coupling of the rotational and vibrational motion of the molecules
with the quantum motion of the solvent is permitted by the similarity 
between energy levels, the electronic structure of an atom is characterized 
by energy differences orders of magnitude larger than the ones needed to 
induce excitation in the atomic motion. 
Although this difference might seem to work in the direction of
simplifying the physical description of the electronic transition processes, 
many important 
details still wait to be clarified. As an example, the fluorescent D$_2$ 
emission line (i.e. the $^2$S$_{1/2}\leftarrow ^2$P$_{3/2}$ radiative 
transition) of heavy single valence electron atoms dispersed in superfluid 
helium is absent,
while the D$_1$ line is sharp and only slightly shifted (1-2 nm) 
to the blue \cite{yabu}. This is in contrast 
with the large broadening and strong blueshift of the absorption lines. 
Moreover, some features of the LIF spectra of the dispersed Ag were interpreted 
as signature of the AgHe and AgHe$_2$ exciplex formation \cite{tak1}.

The blueshift and broadening of the absorption lines
have been interpreted by means of a "bubble model". Here, the dispersed atom
is enclosed in a spherical cavity due to the exchange repulsion 
of its valence electrons and the He ones. 
The liquid He around an atom is modeled
by an isotropic sharp-edge density profile with no atomic internal 
structure. However, both the simple spherical bubble model \cite{kino1}
and the one where quadrupolar distortions of the spherical cavity are allowed 
\cite{kino2} neither quantitatively predict the absorption spectrum of Cs 
and Rb, nor allow to interpret the small splitting of the Rb D$_2$ line.
Reasonably, the lack of any shell structure in the helium density profile,
the absence of a full atomistic description during 
the excitation process, and the physically incomplete description of the bubble
distortion by means of simple quadrupolar deformations
may be held responsible for this undesired outcome \cite{ogata}.

In order to gain a better understanding of the excitation process
and its dependency on the degree of "solvation" of the impurity,
we feel a direct many-body simulation of the excitation spectra to be mandatory.
This also allows to explore the change in the spectra upon the increase of the
number of He atoms in the clusters, and, at the same time, to test the 
validity of our theoretical approach. 

With these goals in mind, we present a
diffusion Monte Carlo study of the $^2\mathrm{P}_{3/2}\leftarrow^2
\mathrm{S}_{1/2}$
and $^2\mathrm{P}_{1/2}\leftarrow^2\mathrm{S}_{1/2}$ absorption spectra of 
silver doped helium clusters. The Ag spectrum, both in bulk helium and in He 
clusters, has been deeply studied and well characterized \cite{tak1,tak2,clo1} 
showing that Ag is indeed solvated.
Moreover, accurate interaction potentials between He and the excited 
$^2$S$_{1/2}$, $^2$P$_{1/2}$, 
and $^2$P$_{3/2}$ states of Ag are available \cite{tak3}. 
These potentials allowed to assign the broad band at 382 nm in the fluorescence 
spectrum to the AgHe$_2$ exciplex ~\cite{tak1}.

To tackle the atomic description needed to compute the excitation spectra, 
we believe the Monte Carlo methods are the best suited techniques.
Since these methods are well 
described in the literature \cite{reybook}, we simply state that while 
variational Monte Carlo (VMC) allows one to optimize a trial wave function 
$\Psi_T(\mathbf{R})$
and to successively compute any expectation values $\langle O \rangle_{VMC}$ 
from it, diffusion Monte Carlo (DMC) corrects the remaining deficiencies of the 
variational description projecting out all the excited state components
sampling $f(\mathbf{R})=\Psi_0(\mathbf{R})\Psi_T(\mathbf{R})$, or less commonly
$f(\mathbf{R})=\Psi_0^2(\mathbf{R})$.

In atomic units, the Hamiltonian operator for our AgHe$_n$ clusters reads as
\begin{equation} 
\label{ageq1} {\mathcal H} = -\frac{1}{2} \left( \sum _{i=1}^{n} 
\frac{\nabla _i ^{2}}{m_{^4 He}} + \frac{\nabla_{\mathrm{Ag}}^{2}}
{m_{\mathrm{Ag}}} \right) + V(\mathbf{R}) 
\end{equation} 
\noindent
Here, we assume a pair potential of the form 
$V(\mathbf{R}) = \sum _{i<j} V_{\mathrm{HeHe}}(r_{ij}) + \sum _i 
V_{\mathrm{AgHe}}(r_i)$ for the clusters with the silver atom in the $^2$S$_{1/2}$
electronic ground state. For $V_{\mathrm{HeHe}}(r_{ij})$ we employed the 
TTY potential \cite{tty}, and for $V_{\mathrm{AgHe}}(r_i)$
we fitted the $^2\Sigma$ AgHe potential by Jakubek and Takami \cite{tak3}.
We computed the energy of the AgHe dimer by means of a grid method \cite{hinze} 
obtaining -4.021 cm$^{-1}$: this value differs from their
result (-4.000 cm$^{-1}$) by only 0.021 cm$^{-1}$. 

Our trial wave function has the common form
$\Psi_T(\mathbf{R})=\prod _{i<j}^N\psi(r_{ij})\prod_i^N\phi (r_{i})$,
where no one-body part was used, and 
$\psi (r) = \phi (r) = exp [ -\frac{p_5}{r^5} -\frac{p_3}{r^3}
-\frac{p_2}{r^2} -p_1 r -p_0 \ln(r)]+a\exp[-b(r-r_0)^2]$.
The parameters of the model wave function were fully
optimized minimizing the variance of the local energy for each cluster. 
The sampled distributions were used to compute exactly the energy using 
the mixed estimator
\begin{equation}
\label{ageq3}
\langle {\mathcal H} \rangle_M = \frac{\int f ({\mathbf R})
{\mathcal H}_{loc} ({\mathbf R}) d{\mathbf R}}
{\int f ({\mathbf R}) d{\mathbf R}}
\end{equation}
\noindent
as well as the mixed and second order estimate $\langle{\mathcal O}\rangle_{SOE}
=2\langle{\mathcal O}\rangle_{M}-\langle{\mathcal O}\rangle_{VMC}$ of many other
expectation values (e.g. the interparticle distribution functions) 
~\cite{reybook}.  

The resulting DMC energy values for the AgHe$_n$ clusters with $n$ up to 100 
are shown in Table \ref{tab1} together with the differential quantity 
$\Delta(n)= -[E(n)-E(m)]/(n-m)$, which can be interpreted as the 
evaporation energy of an He atom from the cluster.
AgHe$_m$ is the largest cluster having $m<n$.
\begin{table}
\begin{center}
\begin{tabular}{lcccc}  \hline\hline
$n$& E($n$) &$\Delta(n)$&D$_1$ &D$_2$ \\ \hline
free Ag &           &            & 338.3 &328.1\\
1  &-4.0212(9) &            &       &     \\
2  &-8.2333(5) &   4.212(1) & 344.5 &334.8\\
3  &-12.598(2) &   4.365(2) & 344.2 &334.6\\
4  &-17.112(1) &   4.514(2) &       &     \\
6  &-26.478(2) &   4.682(2) & 342.6 &333.3\\
8  &-36.259(4) &   4.890(2) & 341.3 &332.3\\
12 &-56.68(1)  &            & 339.6 &329.8\\
13 &-61.78(1)  &   5.09(1)  &       &     \\
14 &-66.84(1)  &   5.06(2)  &       &     \\
15 &-71.61(5)  &   4.77(6)  & 338.3 &328.5\\
19 &-89.31(2)  &            &       &     \\
20 &-93.17(3)  &   3.86(4)  & 337.3 &327.6\\
24 &-107.14(4) &   3.49(1)  &       &     \\
25 &-110.3(1)  &   3.2(1)   &       &     \\
29 &-123.17(7) &            &       &     \\
30 &-126.11(7) &   3.0(1)   & 336.2 &326.6\\
40 &-158.70(6) &   3.26(1)  & 335.5 &325.7\\
50 &-191.3(3)  &   3.27(3)  & 334.7 &324.9\\
60 &-225.1(2)  &   3.38(4)  & 333.8 &323.9\\
70 &-259.9(4)  &   3.47(5)  & 333.4 &323.6\\
80 &-292.4(7)  &   3.26(8)  & 332.5 &322.7\\
90 &-326.2(7)  &   3.38(9)  & 331.9 &322.0\\
100&-357.3(6)  &   3.10(9)  & 331.6 &321.7\\

\hline \hline

\end{tabular}
\caption{Total and evaporation energy (cm$^{-1}$),
D$_1$ and D$_2$ absorption wavelengths (nm) for AgHe$_n$ clusters.}
\label{tab1}
\end{center}
\end{table}

From this data it can be noticed that $\Delta(n)$ does not possess a monotonic
behavior. Instead, the steady increase for $n<13$ is followed by a rapid 
decrease in value before plateauing for $n\sim 25$. This behavior could be
interpreted invoking different effects. For $n<13$, a newly added helium feels
the bare Ag interaction potential plus the interaction with the already present
He atoms, that acts positively increasing the binding energy. 
Quantitatively, we found the changes of $\Delta(n)$ versus $n$ similar to the 
ones obtained for He$_n$ \cite{lew}, He$_n$H$^-$ \cite{heh-}, and 
He$_n$HF \cite{lew_hf}.
Since this effect seems to be independent of the nature of the doping impurity,
one may interpret it as a dynamical many-body effect of the interacting 
helium atoms.

Beyond AgHe$_{13}$, the value of $\Delta(n)$ decreases
indicating the onset of a repulsive interaction. This could be attributed
to an "excluded volume" effect, where each new He is strongly attracted
by Ag in its first coordination shell, but has to "find room" for itself
forcing the other atoms to increase their local density, and rising 
their average kinetic energy.

Finally, for clusters larger than AgHe$_{25}$, the evaporation energy 
remains roughly constant around 3.1-3.5 cm$^{-1}$ indicating that a new He 
atom feels a quite different environment than for $n<25$. 

Figure 1 represents the interparticle 
Ag-He probability density functions for clusters having $n$ from 12 to 100.  
\begin{figure}
\label{fig2}
\rotatebox{-90}{
\includegraphics{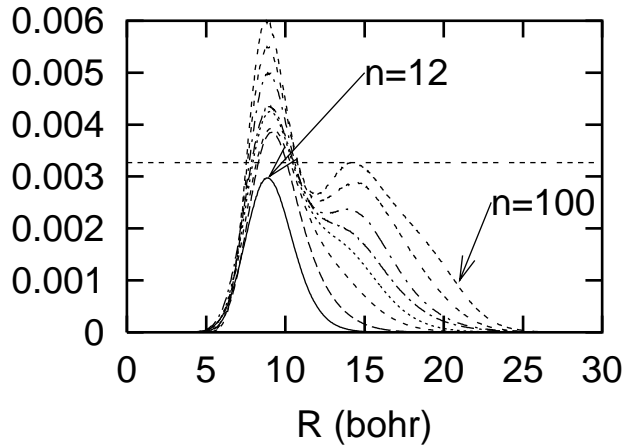}
}
\caption{He density distributions around Ag
for $n=$12, 20, 30, 40, 50, 60, 80, and 100.}
\end{figure}
These were normalized so that
$4\pi\int_0^{\infty} r^2 \rho(r) dr = n$, therefore representing the local 
density of He atoms around Ag.  The functions for $n<12$ 
overlap in shape with the AgHe$_{12}$ 
one. As to AgHe$_{30}$, the presence of a broad shoulder at large $r$, 
that successively develops into a well defined peak, unambiguously indicates 
a second shell. More interestingly, the height of the first peak 
continuously rises until the second shell is completely filled, as indicated by 
the onset of another shoulder at large Ag-He distance for AgHe$_{100}$.
Moreover, the density minimum between the first and second shell peaks also
increases in height on going towards larger clusters, becoming just 15\%
less than the second shell peak height. Both these evidences can be
interpreted as a direct signature of the "non rigidity" of the first He layer,
as well as of an easy exchange process between the first and second shell
\cite{reatto_ion}.

As to the angular distributions, Figure 2 shows several
$\cos(\mathrm{HeAgHe})$ distributions.
\begin{figure}
\label{fig3}
\rotatebox{-90}{
\includegraphics{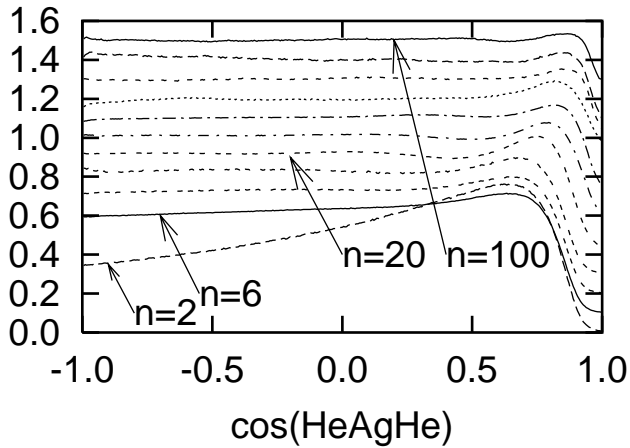}
}
\caption{$\cos(\mathrm{HeAgHe})$ density distributions for 
$n=$2, 6, 8, 12, 20, 30, 40, 50, 60, 80, and 100. Each distribution is shifted 
upwards with respect to the previous one by 0.1.}
\end{figure}
The smaller clusters ($n\leq20$) show a
deep minimum for $\cos(\mathrm{HeAgHe})=1$ and a smooth maximum
located in the 0.6-0.8 range, both strong indications
of a structured distribution of the He atoms in the first solvation
shell. Here, the minimum indicates that two He atoms cannot 
overlap or surmount each other, the only two possible arrangements having
$\cos(\mathrm{HeAgHe})=1$. Whereas the overlap is forbidden by the 
repulsive part of the He-He potential, the possibility of an He atom to surmount
another is hindered by the strength of the $^2\Sigma$ AgHe potential
that forces the He motion in a limited radial region around Ag as shown
by the AgHe$_{12}$ radial distribution. Instead, the smooth maximum indicates 
the relative localization effects due to the attractive interaction between 
He atoms.
This effect is particularly evident for AgHe$_{2}$, whose angular distribution
function decreases on going towards $\cos(\mathrm{HeAgHe})=-1$.
The position of the maximum shifts to larger
$\cos(\mathrm{HeAgHe})$ values on going from $n=2$ to $n=20$,
suggesting a progressively more structured packing of the He atoms in the
first shell for $n=20$, and agreeing nicely with the aforementioned "excluded
volume" interpretation. The structured packing is also supported by the 
shallow second peak located around 0.1 in the AgHe$_{20}$ cosine distribution.
Both the minimum and the maximum are "smeared out" by adding He atoms to 
AgHe$_{20}$, a clear indication that the second shell is less structured and 
more fluid than the first one.
 
As to the absorption spectrum of the embedded Ag atom, we computed this
observable using the semiclassical approach proposed by Lax \cite{lax}, and 
adapted to the quantum Monte Carlo framework by Cheng and Whaley \cite{cheng}.
In this method, 
the spectral lines are computed collecting the distribution of the difference
$V_{exc}(\mathbf{R})-V_{gs}(\mathbf{R})$ or, more accurately, of the quantity
$V_{exc}(\mathbf{R})+\sum _{i<j} V_{\mathrm{HeHe}}(r_{ij})-E_0$ over the sampled
$f(\mathbf{R})$ \cite{maxcom}. 
Here, $V_{gs}(\mathbf{R})$ ($V_{exc}(\mathbf{R})$) is 
the interaction potential between the ground (excited) state Ag atom with 
the sourronding He atoms, while $E_0$ is the DMC ground state energy.
The three $V_{exc}(\mathbf{R})$ PES for a given cluster configuration are
obtained from the AgHe $^2\Pi_{1/2}$, $^2\Pi_{3/2}$, and $^2\Sigma$ 
interaction potentials using the Diatomic-in-Molecules approach. 
The details are well
described by Nakayama and Yamashita for the Li, Na, and K cases \cite{yama}.

\begin{figure}
\label{fig4}
\rotatebox{-90}{
\includegraphics{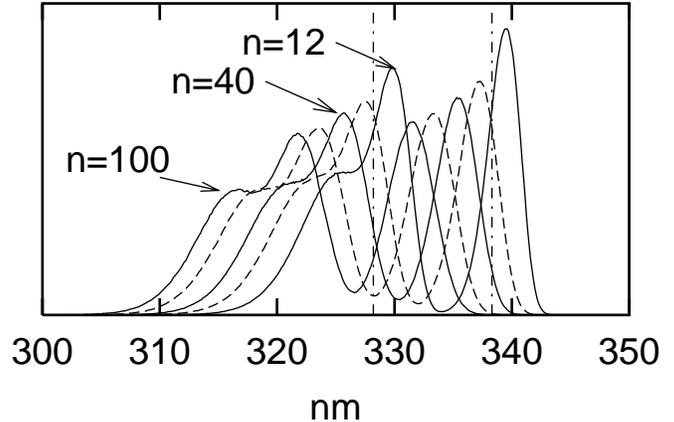}
}
\caption{Simulated absorption spectra for AgHe$_n$ clusters with 
$n=$12, 20, 40, 70, and 100. The vertical lines represent the free Ag spectrum.}
\end{figure}
The spectra obtained collecting $V_{exc}(\mathbf{R})-V_{gs}(\mathbf{R})$ 
during the simulations are shown in Table \ref{tab1}, and in Figure 3
for several representative clusters. The same quantities obtained by collecting
$V_{exc}(\mathbf{R})+\sum _{i<j} V_{\mathrm{HeHe}}(r_{ij})-E_0$ 
are blueshifted by less than 1 nm. The computed spectra clearly show the two 
separated bands deriving from the excitation of Ag into $^2P_{1/2}$ 
and $^2P_{3/2}$ states, 
the second one also displaying the classical short wavelength shoulder typical 
of the D$_2$ line of heavy alkali atoms in superfluid helium \cite{kino1,kino2}.
For our largest cluster, the D$_1$ and D$_2$ lines have maxima located at 
331.6 and 321.7 nm, and a FWHM of 4.3 and 9.8 nm, respectively. These results
are in accurate agreement with the experimental wavelengths
332.8 and 322.5 nm, and FWHM 4.0 and 8.5 nm ~\cite{clo1}. 

From the spectra shown in Fig. 3, it clearly appears that the broadening 
of the absorption bands increases on going towards larger clusters. 
This evidence indicates that
the Ag electronic degrees of freedom are coupled with the motion of an 
increasing number of He atoms, and not only with those located in the first 
shell. More interestingly, whereas all the clusters with $n\leq 15$ show a 
redshift with respect to the free Ag lines, the ones with $n\geq19$ display a 
blueshift strongly dependent on the number of He atoms. 
Here, the redshift for $n\leq 15$ indicates that the clusters possess 
an internal distribution such that a vertical transition brings them in a
region of the excited state potential where the complexes can form a bound 
state. This may give the possibility of producing AgHe$_n$ ($n=$1-15)
exciplexes starting from the corresponding clusters, and to experimentally
study their spectrum and decaying dynamics. Conversely, the larger clusters
are vertically excited to repulsive regions of the potential energy surface 
(PES), therefore preventing
the direct formation of larger exciplexes.

The blueshift for $n\geq19$, at variance with basic solvation concepts,
indicates a large effect of the second shell filling on the absorption 
wavelengths. This is confirmed by the computational evidence that the
excitation spectrum of AgHe$_{100}$, that shows the onset of a third shell,
closely agrees with the one of AgHe$_{90}$ (see Table \ref{tab1}).

In order to rationalize this observation, as well as the monotonic blueshift 
of the absorption bands upon increasing of $n$, one must notice 
that the portion of the AgHe pair distribution located in the 10-13 bohr 
range overlaps with the tail of the repulsive excited AgHe $^2\Sigma$ potential.
As a consequence, this zone of the pair density introduces some positive 
contribution to the diagonal elements of the matrix whose eigenvalues define 
the three electronic excited PES of the complexes. 
Since the magnitude of these contributions is dependent on the local He density
via the sum $\sum V_{^2\Sigma}(r)$ over the He atoms falling in that range,
there is a net increase of the values of the diagonal elements upon increasing
of the size of the cluster. This fact reflects itself in a positive
shift of the eigenvalues, and hence in the blueshift of the computed spectra.

The authors thank Prof. Michio Takami for sending the computed interaction 
potentials.  This work was supported by Italian MIUR Grant No. MM03265212.
The authors are indebted to the
Centro CNR per lo Studio delle Relazioni tra Struttura e Reattivita' Chimica 
for grants of computer time.


\begin{thebibliography}{99}
\bibitem{maxcom} While Cheng and Whaley \cite{cheng} were forced to approximately estimate the contribution from the kinetic energy of the doping impurity to the computed solid matrix spectra, for finite systems as our clusters this can be exactly accounted for simply using the total ground state energy.
\end{thebibliography}
\end{document}